\documentclass[%
 aip,
 rmp,
 reprint,
 superscriptaddress,
 longbibliography,
 noamsfonts,		
 noamssymb,
 noamsmath,		
 preprintnumbers,
 floatfix,
 showpacs,
 byrevtex,
]{revtex4-1}

\usepackage{comment}
\usepackage{microtype}

\usepackage{color}
\usepackage{ifpdf}
 \ifpdf
  \usepackage[pdftex]{graphicx}
 \else
  \usepackage{graphicx}
\fi
\usepackage{color}

\usepackage[UKenglish]{babel}

\usepackage[autoload,graphics]{svn-multi}
 \svnid{$Id: main.tex 20 2013-02-14 13:04:38Z bt $}
 \svnidlong
 {$HeadURL$}
 {$LastChangedDate: 2013-02-14 14:04:38 +0100 (Thu, 14 Feb 2013) $}
 {$LastChangedRevision: 20 $}
 {$LastChangedBy: bt $}

\usepackage[T1]{fontenc}
\usepackage[scaled=0.92]{helvet}

\usepackage{mathptmx}
\usepackage{mathtools}
\usepackage{bm}
\usepackage{siunitx}

\RequirePackage{xspace}
\newcommand*{\cf}{\emph{cf.}\xspace}

\newcommand*{\ie}{\emph{i.e.}\xspace}

\newcommand*{\D}[1]{\mathrm{d}#1}
\newcommand*{\epsz}{\varepsilon_{0}}
\newcommand*{\im}{{\mathrm{i}}}

\newcommand*{\Sup}[1]{^{\text{#1}}}
\newcommand*{\w}{\omega}

\providecommand{\mbf}{\mathbf}
\providecommand{\hm}{\bm}
\renewcommand*{\vec}[1]{\mbf{#1}}
\newcommand*{\tens}[1]{\pmb{\mathsf{#1}}}
\newcommand*{\bdot}{\hm{\cdot}}

\newcommand*{\del}{\hm{\nabla}}
\renewcommand*{\div}{\del\bdot}
\newcommand*{\cross}{\bm{\times}}
\newcommand*{\unit}[1]{\hat{\vec{#1}}}
\newcommand*{\nunit}{\unit{n}}
\newcommand*{\0}{\vec{0}}
\newcommand*{\B}{\vec{B}}
\newcommand*{\E}{\vec{E}}
\newcommand*{\F}{\vec{F}}
\newcommand*{\g}{\vec{g}}
\newcommand*{\gfield}{\g\Sup{field}}
\newcommand*{\gmech}{\g\Sup{mech}}
\newcommand*{\h}{\vec{h}}
\newcommand*{\hfield}{\h\Sup{field}}
\newcommand*{\hmech}{\h\Sup{mech}}
\renewcommand*{\j}{\vec{j}}
\newcommand*{\J}{\vec{J}}
\newcommand*{\Jfield}{\J\Sup{field}}
\newcommand*{\Jmech}{\J\Sup{mech}}
\newcommand*{\veck}{\vec{k}}
\newcommand*{\p}{\vec{p}}
\newcommand*{\pfield}{\p\Sup{field}}
\newcommand*{\pmech}{\p\Sup{mech}}
\newcommand*{\vtau}{\boldsymbol{\tau}}
\newcommand*{\x}{\vec{x}}

\renewcommand*{\D}[1]{\mathrm{d}#1}
\newcommand*{\dV}{\D{^3}\!x}
\newcommand*{\dd}[2]{\frac{\D{#2}}{\D{#1}}}
\newcommand*{\pdd}[2]{\frac{\partial{#2}}{\partial{#1}}}
\newcommand*{\dS}{\D{^2}\!x}
\newcommand*{\dSvec}[1]{\dS{#1}\,\nunit{#1}}
\newcommand*{\intV}[2]{\int_{V{#1}}\!\dV{#1}\,{#2}}
\newcommand*{\ointSvecdot}[2]{\oint_{S{#1}}\!\dSvec{#1}\bdot{#2}}

\newcommand*{\ARI}{%
 ARI, Sezione Venezia,
 Santa Croce~1996,
 IT-30123 Venice, Italy, EU%
}

\newcommand*{\GHT}{%
 GHT Photonics~srl,
 via Istria~55,
 IT-35135 Padova, Italy, EU%
}
\newcommand*{\IRF}{%
 Swedish Institute of Space Physics,
 {\AA}ngstr\"{o}m Laboratory,
 P.\,O.~Box~537,
 SE-75121 Uppsala, Sweden, EU%
}
\newcommand*{\LaNN}{%
 LaNN Venetonanotech,
 via Stati Uniti~4,
 IT-35100 Padova, Italy, EU%
}
\newcommand*{\ROVER}{%
 Ro.ver Instruments,
 via Parini~2,
 IT-25019 Sirmione (VR), Italy, EU%
}
\newcommand*{\TWISTOFF}{%
 Twist-Off~srl,
 via Croce Rossa 112,
 IT-35129 Padova, Italy, EU%
}
\newcommand*{\UNIPD}{%
 Department of Physics and Astronomy,
 University of Padova%
}
\newcommand*{\UNIPDviaMar}{%
 \UNIPD,
 via Marzolo~8,
 IT-35131 Padova, Italy, EU%
}
\newcommand*{\UNIPDviaOss}{%
 \UNIPD,
 vicolo dell'Osservatorio~3,
 IT-35122 Padova, Italy, EU%
}

\begin{document}

\preprint{SVN version \svnauthor-\svnrev}

\title{Experimental demonstration of free-space information transfer using
 phase modulated orbital angular momentum radio}

\author{F. Tamburini} 
\altaffiliation[Also at: ]{\ARI}
\affiliation{\TWISTOFF}

\author{B.\,Thid\'e}
\email[E-mail: ]{bt@irfu.se}
\affiliation{\IRF}

\author{V. Boaga} 
\affiliation{\ARI}

\author{F. Carraro} 
\altaffiliation[Also at: ]{\ROVER}
\affiliation{\ARI}

\author{M. del Pup} 
\affiliation{\ARI}

\author{A. Bianchini} 
\affiliation{\UNIPDviaOss}

\author{C.\,G. Someda} 
\affiliation{\GHT}

\author{F. Romanato} 
\altaffiliation[Also at: ]{\LaNN}
\affiliation{\UNIPDviaMar}

\date{\svndate}

\begin{abstract}

In a series of fundamental proof-of-principle studies, including
numerical, controlled indoor laboratory, and real-world outdoor
experiments, we have shown that it is possible to use electromagnetic
angular momentum as a physical layer for radio science and radio
communication applications
\cite{Thide&al:PRL:2007,%
Tamburini&al:NP:2011,%
Tamburini&al:APL:2011,%
Tamburini&al:NJP:2012,%
Tamburini&al:NJP:2012a%
}.
Here we report a major, decisive step toward the realization of the
latter, in the form of a real-world experimental demonstration that
a radio beam carrying orbital angular momentum (OAM) can readily be
digitally phase shift modulated and that the information thus encoded
can be effectively transferred in free space to a remote receiver.
The experiment was carried out in an urban setting and showed that
the information transfer is robust against ground reflections and
interfering radio signals. The importance of our results lies in the
fact that digital phase shift keying (PSK) protocols are used in many
present-day wireless communication scenarios, allowing new angular
momentum radio implementations to use methods and protocols that are
backward compatible with existing linear momentum ones.

\end{abstract}


\maketitle

\section{Introduction}

Current radio science and communication implementations based on the
electromagnetic (EM) linear momentum (Poynting vector) physics layer are
beginning to approach their limits in terms of radio frequency spectrum
availability and occupancy. This calls for the introduction of 
new radio paradigms. For this purpose, it has been proposed that the EM
angular momentum, well described in the standard literature
\cite{%
Heitler:InbookAppendix1:1954,%
Bogolyubov&Shirkov:InbookChap2:1959,%
Messiah:InbookChapXXI:1970,%
Eyges:InbookChap11:1972,%
Berestetskii&al:InbookChap1:1989,%
Ribaric&Sustersic:Book:1990,%
Mandel&Wolf:Inbook:1995,%
Cohen-Tannoudji&al:Inbook:1997,%
Schwinger&al:Inbook:1998,%
Jackson:InbookChap7&12:1998,%
Allen&al:Book:2003,%
Rohrlich:Inbook:2007,%
Andrews:Book:2008,%
Torres&Torner:Book:2011,%
Yao&Padgett:AOP:2011,%
Thide:Inbook:2011%
}
but hitherto underutilized in radio science and technology,
be fully exploited in radio communications
\cite{%
Thide&al:PRL:2007,%
Franke-Arnold&al:LPR:2008,%
Thide&al:Incollection:2011,%
Tamburini&al:NP:2011,%
Tamburini&al:APL:2011,%
Tamburini&al:NJP:2012,%
Tamburini&al:NJP:2012a%
}.
As has been amply demonstrated in experiments at optical frequencies,
the use of EM angular momentum can indeed increase the information
entropy and hence the capacity of wireless communications
\cite{%
Gibson&al:OE:2004,%
Celechovsky&Bouchal:NJP:2007,%
Barreiro&al:NP:2008,%
Pors&al:PRL:2008,%
Martelli&al:EL:2011,%
Kumar&al:OL:2011,%
Wang&al:NPHO:2012%
}.
This is consistent with the fact that all classical fields carry angular
momentum \cite{Belinfante:Physica:1940,Soper:InbookChap9:1976} and that
the dynamics of a general physical system is not fully described unless
both its total linear momentum and total angular momentum are specified
\cite{Truesdell:InbookChap5:1968}. Specifying only one of them is not
sufficient. And not using both the linear and the angular momentum of an
EM field is therefore not using the field to its full capacity.

In order to add information transfer capacity to present-day radio
communication links, it is common practice to invoke EM spin angular
momentum (SAM) in the form of wave polarization. SAM is an intrinsic
property of the classical EM field (and each individual photon),
describing the spin characteristics of the EM rotational degrees
of freedom. However, in free space SAM can attain only two values,
$\sigma=\pm1$, corresponding to left-hand and right-hand wave
polarization, respectively. In other words, SAM spans a two-dimensional
state space (Hilbert space). By invoking SAM one can therefore, at
most, double the information transfer capacity within a given frequency
bandwidth. The physical encoding of a beam with SAM causes the EM field
phases of the beam to be different in \emph{different directions at
one and the same point} in a plane perpendicular to the beam axis. A
common technique to demonstrate and exploit SAM in radio is to use two
co-located orthogonal dipole antennas, also known as turnstile antennas.
Another common technique is to use helical antennas.

\begin{table*}
\caption{\label{tab:g_and_h}%
 Similarities and differences between the EM linear momentum
 density $\gfield(t,\x)$ and the EM angular momentum density
 $\hfield(t,\x,\x_0)$ around a moment point $\x_0$ carried by a
 classical electromagnetic field $[\E(t,\x),\B(t,\x)]$ in the
 presence of matter (particles) with mechanical linear momentum
 density $\gmech(t,\x)$ and mechanical angular momentum density
 $\hmech(t,\x,\x_0)$.%
}
\begin{ruledtabular}
\begin{tabular}{lcc}
  \textbf{Property}
 &
  \textbf{Linear momentum density}
 &
  \textbf{Angular momentum density}
\\
 Definition 
 &
  $\gfield=\epsz\E\cross\B$
 &
  $\hfield=(\x-\x_0)\cross(\epsz\E\cross\B)$
\\
 SI unit
 &
  \si{N.s.m^{-3}} (\si{kg.m^{-2}.s^{-1}})
 &
  \si{N.s.m^{-2}} (\si{kg.m^{-1}.s^{-1}})
\\
 Spatial fall off at large distances $r$
 &
  $\sim r^{-2} + \mathcal{O}(r^{-3})$ 
 &
  $\sim r^{-2} + \mathcal{O}(r^{-3})$ 
\\
 Typical phase factor
 &
  $\exp{\{\im(\veck\bdot\x - \w t)\}}$
 &
  $\exp{\{\im(\veck\bdot\x -\w t + \alpha\varphi)\}}$ 
  , $0\le\varphi<2\pi$
\\[.5ex]
 Local conservation law 
 &
  $\pdd{t}{\gmech} + \pdd{t}{\gfield} + \div\tens{T} = \0$
 &
  $\pdd{t}{\hmech} + \pdd{t}{\hfield} + \div\tens{K} = \0$
\\[1ex]
 C (charge conjugation) symmetry
 &
 Even 
 &
 Even
\\
 P (spatial inversion) symmetry
 &
 Even (polar vector, ordinary vector)
 &
 Odd (axial vector, pseudovector)
\\
 T (time reversal) symmetry
 &
 Odd
 &
 Odd
\end{tabular}
\end{ruledtabular}
\end{table*}

In contrast, the EM orbital angular momentum (OAM) is an extrinsic
property of the EM field (and each individual photon), describing the
orbital characteristics of its rotational degrees of freedom. Associated
with OAM is an EM field phase factor $\exp{\{\im\alpha\varphi\}}$
where $\varphi$ is the azimuthal angle around the beam axis. Of course,
the phase function $\alpha$ may attain any value, not necessarily
only integers. However, due to the single-valuedness of the
fields, OAM is quantized such that an EM field that carries
non-integer OAM is a weighted superposition of discrete OAM eigenmode
components, each of which is proportional to $\exp{\{\pm\im
m\varphi\}}$, where $m=0,1,2,\ldots$, acts as a quantum number
\cite{Berry:JOA:2004c,Tamburini&al:APL:2011}, also called the
topological charge.

Hence, a beam carrying an arbitrary amount of OAM contains a spectrum
of discrete integer OAM eigenmodes \cite{Torner&al:OE:2005} that are
mutually orthogonal (in a function space sense) and therefore propagate
independently \cite{Molina-Terriza&al:PRL:2002}. In other words,
OAM spans a state space of dimensionality $N=1,2,3,\ldots$, and can
therefore be regarded as an `azimuthal polarization' with arbitrarily many
states \cite{Litchinitser:Science:2012}. This makes it possible, at
least in principle, to use OAM to physically encode an unlimited amount
of information onto any part of an EM beam, down to the individual
photon \cite{Mair&al:N:2001}. The physical encoding of a beam with
OAM causes the EM field of the beam to attain a whole set of unique
characteristics. One of these characteristics is that the
EM field exhibits different phases when measured in \emph{one and the
same direction at different points} in a plane perpendicular to the
beam axis. In our experiments we can therefore use standard phase
interferometers constructed from ordinary (linear momentum) antennas to
analyze the OAM content of the beams.

It should be emphasized that for EM beams of the kind used in the
experiment described here, OAM is distinctively different from---and
independent of---SAM (wave polarization). If such a beam is already
$N$\!-fold OAM encoded, adding SAM will double the information transfer
capacity by virtue of the fact that the dimensionality of the state
space doubles from $N$ to $2N$. So far we have not utilized the
polarization (SAM) degree of freedom in our OAM radio experiments. Such
experiments are pending.

Proof-of-concept studies have shown that it is possible to use the
total angular momentum, \ie\ SAM+OAM, as a new physical layer for radio
science and technology exploitation. These studies include numerical
experiments \cite{Thide&al:PRL:2007} showing that it is feasible to
utilize OAM in radio; controlled anechoic chamber laboratory experiments
\cite{Tamburini&al:APL:2011} verifying that it is possible to generate
and transmit radio beams carrying non-integer OAM and to measure their
OAM spectra in the form of weighted superpositions of different integer
OAM eigenstates; and outdoor experiments \cite{Tamburini&al:NJP:2012}
verifying that in a real-world setting different signals, encoded in
different OAM states, can be transmitted independently to a receiver
located in the (linear momentum) far zone and be resolved there.

We have therefore proposed that the angular momentum physical layer
be used as an alternative and/or supplement to the linear momentum
(Poynting vector) physical layer that is used in current radio
communication implementations. As an important step in the practical
implementation of this, we recently demonstrated experimentally that
information encoded in OAM radio beams in terms of quadrature phase
shift coding (QPSK) modulation can be robustly transferred in free space
even in the presence of reflections and interfering radio signals. The
results of this experiment are presented here.

\section{Physical background}

Let $\gfield$ and $\hfield$ denote the volumetric EM linear and angular
momenta densities, respectively, some properties of which are listed in
Table~\ref{tab:g_and_h}. Then the total classical EM linear momentum,
localized inside a volume $V$ in free space where the dielectric
permittivity is $\epsz$, is given by
\cite{
Schwinger&al:Inbook:1998,%
Rohrlich:Inbook:2007,%
Thide&al:Incollection:2011%
}
\begin{subequations}
\label{eq:EM_momenta}
\begin{align}
\begin{split}
\pfield(t)
 &= \intV{}{\gfield(t,\x)}
\\
 &= \epsz\intV{}{[\E(t,\x)\cross\B(t,\x)]} 
\end{split}
\end{align}
and the total classical EM angular momentum (SAM+OAM) around an arbitrary
moment point $\x_0$ carried by the same EM field in this volume $V$ is
given by
\cite{
Schwinger&al:Inbook:1998,%
Rohrlich:Inbook:2007,%
Thide&al:Incollection:2011%
}
\begin{align}
\begin{split}
 \Jfield(t,\x_0) 
 &= \intV{}{\hfield(t,\x,\x_0)}
\\
 &= \intV{}(\x-\x_0)\cross\gfield(t,\x)
\\
 &= \epsz\intV{}{(\x-\x_0)\cross[\E(t,\x)\cross\B(t,\x)]}
\\
 &= \Jfield(t,\0) -\x_0\pfield(t)
\end{split}
\end{align}
\end{subequations}

For a localized source, it is convenient to evaluate the integrals in
eqns.~\eqref{eq:EM_momenta} in a spherical polar coordinate system
$(r,\theta,\varphi)$ with its origin at the barycentre of the source
region. A signal pulse emitted by the source during a finite time
interval $\Delta{t}$, will, after it has left the source region, be
propagating radially outward, in the surrounding free space, with speed
$c$ and be confined to a finite volume $V_0$ between two spherical
shells, one with radius $r_0$ relative to the source, and another
with radius $r_0+\Delta{}r_0$ where $\Delta{r_0}=c\Delta{t}$ and
with a certain distribution in the angular $(\theta,\varphi)$ domain
\cite{Schwinger&al:Inbook:1998}. Consequently, the total linear momentum
carried by such an EM pulse propagating in free space is
\begin{multline}
\label{eq:spherical_int}
 \intV{_0}{\gfield(t,\x)}
\\
 = \int_{r_0}^{r_0+\Delta{r_0}}\D{r}\,r^2
   \int_0^{2\pi}\D{\varphi}
   \int_0^\pi\D{\theta}\sin\theta\,{\gfield(t,r,\theta,\varphi)}
\end{multline}

Integration of $\gfield$ over the angular domain [the two last
integrals in eqn.~\eqref{eq:spherical_int}] yields a function of
$r$ (and $t$) that, for very large $r_0$, becomes proportional to
$r^{-2}$ ; \cf\ Table~\ref{tab:g_and_h} and
Refs.~\onlinecite{%
Abraham:PZ:1914,%
Thide&al:ARXIV:2010,%
Tamburini&al:NJP:2012a%
}).
Taking into account that this function shall in the remaining integral
in eqn.~\eqref{eq:spherical_int} be multiplied by $r^2$ and then integrated
over a finite radial interval $[r_0,r_0+\Delta{r_0}]$, the entire
integral, and hence $\pfield$, tends to a constant when $r_0$ tends
to infinity. This asymptotic independence of $r$, allowing the EM
linear momentum generated by a localized source to be transported
all the way to infinity without radial fall off and therefore be
irreversibly lost from the source, is the famous arrow of radiation
asymmetry (see Ref.~\onlinecite[pp.~328--329]{Eddington:Book:2010}, and
Ref.~\onlinecite[Chap.~6]{Jackson:Book:1998}).

Recalling the fact that the angular momentum density $\hfield$ has
precisely the same asymptotic $r^{-2}$ radial fall off as the linear
momentum density $\gfield$ (see Table~\ref{tab:g_and_h}, and
Refs.~\onlinecite{%
Abraham:PZ:1914,%
Thide&al:ARXIV:2010,%
Tamburini&al:NJP:2012a%
}),
it is clear that also the emitted angular momentum tends asymptotically
to a constant at very large distances from the source and is
irreversibly lost there. This is the angular momentum analogue of the
above mentioned linear momentum arrow of radiation.

Consequently, both linear and angular momenta can propagate---and be
used for information transfer---over, in principle, arbitrarily long
distances. Of course, the magnitude and angular distribution of the
respective momentum densities depend on the specific spatio-temporal
and topological EM properties of the actual radiating device used.
Some devices, such as the linear antennas used in radio today, are
effective radiators and sensors of linear momentum whereas angular
momentum is more optimally radiated and sensed by other devices. In
radio engineering parlance, the angular distribution of the linear
momentum density (Poynting vector) is often referred to as the
`radiation pattern' or `antenna diagram'. It should be noted that
for one and the same radiating system this angular distribution
of the \emph{linear} momentum density (`radiation pattern') is
\emph{not} the same as the angular distribution of the \emph{angular}
momentum density. See Refs.~\onlinecite{Thide:Inbook:2011} and
\onlinecite{Then&Thide:ARXIV:2008}.

As shown by well-known conservation laws that follow directly from
Maxwell's equations, the two physical observables total linear momentum
and total angular momentum are conserved (constants of motion). Hence,
$\pfield$ and $\Jfield$ in a fixed volume an EM beam propagating in free
space can neither decrease nor increase.

We recall that the linear momentum fulfils the conservation law
\cite{%
Schwinger&al:Inbook:1998,%
Jackson:InbookChap7&12:1998,%
Thide:Inbook:2011%
}
\begin{subequations}
\label{eq:conservation_laws_global}
\begin{gather}
\label{eq:conservation_law_lin_mom}
 \dd{t}{\pfield} + \F + \ointSvecdot{}{\tens{T}} = \0
\intertext{where $\tens{T}$ is the EM linear momentum flux tensor
(the negative of Maxwell's stress tensor), and}
\label{eq:force}
 \F = \dd{t}{\pmech} 
\end{gather}
\end{subequations}
is the mechanical force (Newton's second law, Euler's first law) on the
particles ($\pmech$ being the mechanical momentum),

The angular momentum fulfils the conservation law
\cite{%
Schwinger&al:Inbook:1998,%
Jackson:InbookChap7&12:1998,%
Thide:Inbook:2011%
}
\begin{subequations}
\label{eq:conservation_law_ang_mom}
\begin{gather}
 \dd{t}{\Jfield(\x_0)} + \vtau(\x_0) + \ointSvecdot{}{\tens{K}(\x_0)}
  = \0
\intertext{where $\tens{K}$ is the EM angular momentum flux tensor and}
\label{eq:torque}
 \vtau(\x_0) = \dd{t}{\Jmech(\x_0)}
\end{gather}
\end{subequations}
is the mechanical torque (Euler's second law) on the particles
($\Jmech$ being the mechanical angular momentum).

Eqns.~\eqref{eq:conservation_law_ang_mom} clearly show that
angular-momentum radio beams should ideally be radiated and sensed by
rotational dynamics devices, \ie, `antennas' based on torque
\cite{%
Beth:PR:1935,%
*Beth:PR:1936,%
Holbourn:N:1936,%
Carrara:N:1949,%
Allen:AJP:1966,%
Carusotto&al:NC:1968,%
Chang&Lee:JOSAB:1985,%
Vulfson:USP:1987,%
Kristensen&al:OC:1994,%
He&al:PRL:1995,%
Friese&al:PRA:1996,%
Then&Thide:ARXIV:2008,%
Helmerson&al:Topologica:2009,%
Padgett&Bowman:NP:2011,%
Ramanathan&al:PRL:2011,%
Elias:AA:2012%
}
rather than on force (translational oscillations of charges, antenna
currents). However, `antennas' for the radio frequency range based on
rotational degrees of freedom and torque are not yet readily available.
On the other hand it has been shown \cite{Thide&al:PRL:2007} that
it is possible to use arrays consisting of a sufficient number of
antennas of the conventional translational (conduction) degree of
freedom type, to generate approximate OAM eigenmodes and superpositions
thereof. Alternatively, one can combine such antennas with reflectors
or lenses that have azimuthally dependent reflective properties
\cite{Tamburini&al:APL:2011}, including helicoidal parabolic antennas
\cite{Tamburini&al:NJP:2012}. This makes it possible already now
to use readily available analogue and digital radio techniques and
technologies, including ordinary linear-momentum sensing and generating
antennas, to study certain fundamental properties of OAM experimentally
in the radio regime \cite{Thide&al:PRL:2007}.

\section{Experimental results}

\begin{figure*}
\centering
 \resizebox{1.\textwidth}{!}{%
  \input{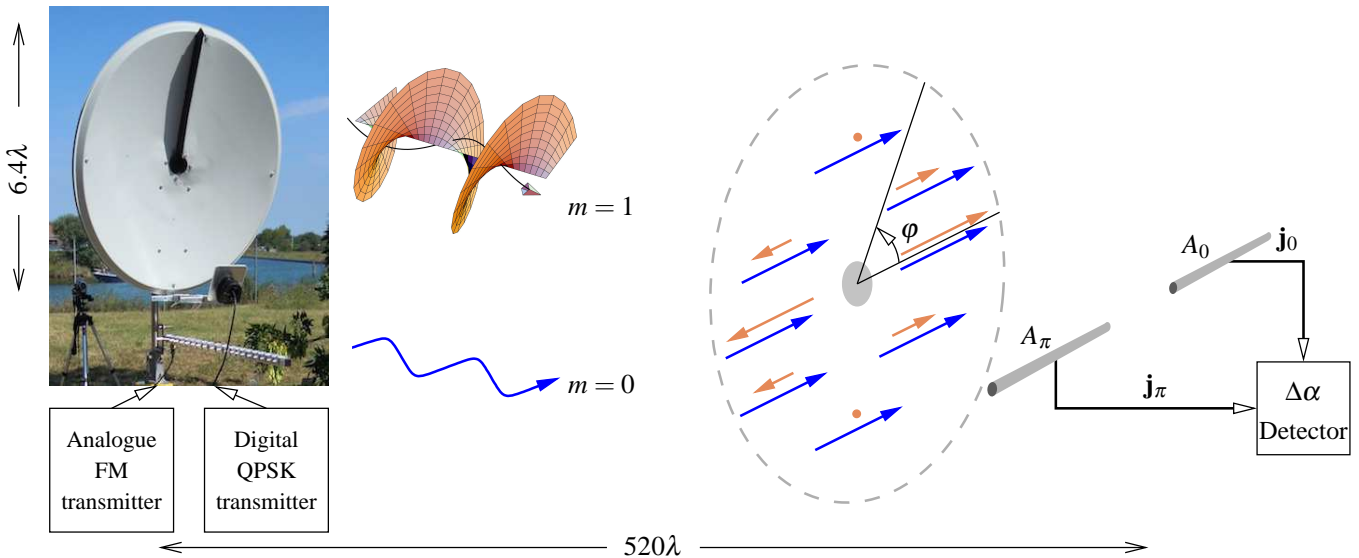}%
 }
\caption{\label{fig:setup}%
 Schematic description of the experimental setup. The OAM $m=1$
 state (colour coded orange) was produced by a helicoidally deformed
 parabolic antenna of the same type as used in the experiment reported
 in Ref.~\onlinecite{Tamburini&al:NJP:2012}. This antenna was fed
 by a digital QPSK transmitter operating at a carrier frequency
 of $2.414$~GHz, transmitting live TV pictures. Horizontal linear
 polarization was used. The linear-momentum ($m=0$) signal at the
 same carrier frequency and with the same polarization (colour coded
 blue) was emitted by a standard Yagi-Uda antenna, fed by an FM
 signal modulated with a TV test pattern. At the receiving end, at a
 distance of $520\lambda$ from the transmitting antennas, well into
 their (linear-momentum) far zones, the instantaneous EM fields of the
 $m=0$ and $m=1$ modes should be in phase on one side of the central
 axis (azimuthal angle $\varphi=0$) and in anti-phase at the opposite
 side ($\varphi=\pi$). Therefore a simple standard linear-momentum
 interferometer was used in the experiment to discern between the
 two OAM eigenmodes. A typical phase interferometer consists of two
 identical antennas $A_0$ and $A_\pi$ producing antenna current
 densities $\j_0$ and $\j_\pi$, respectively. The phase differences
 of these currents are measured and from these measurements the
 differences between the phases of the two electric field vectors are
 estimated. The inset illustrating the twisted beam was taken from
 {www.gla.ac.uk/schools/physics/research/groups/optics}.%
}
\end{figure*}

\begin{figure}[h]
\vspace{1pt}
\centering
 \resizebox{1.\columnwidth}{!}{%
  \input{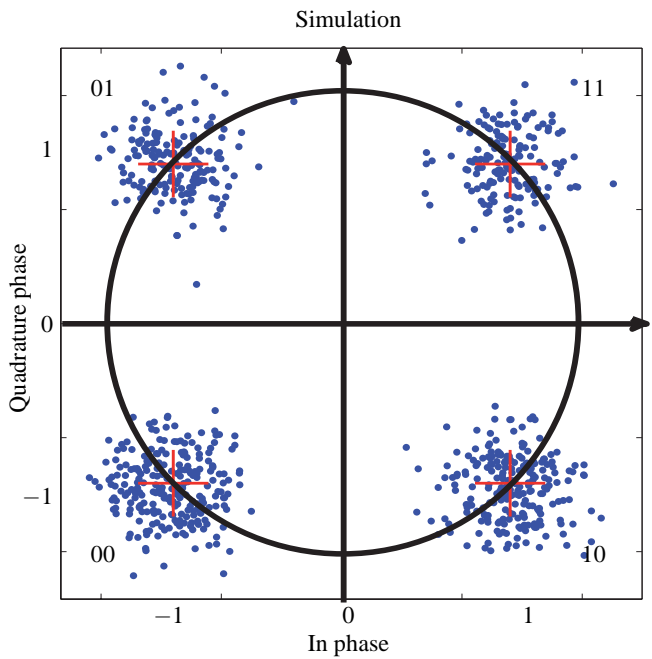}
 }
\caption{%
\label{fig:QPSK_simulated}
 Numerically simulated constellation diagram for the QPSK information
 transfer.%
}
\end{figure}
\begin{figure}[h]
\centering
 \resizebox{1.\columnwidth}{!}{%
  \input{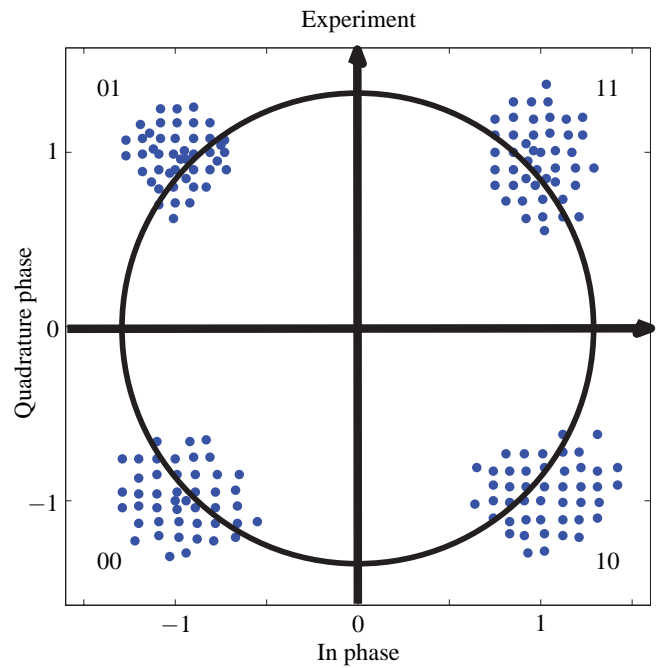}
 }
\caption{%
\label{fig:QPSK_measured}
 Constellation diagram for the QPSK information transfer as measured
 in the experiment in the presence of ground reflections and an interfering
 FM signal at the same carrier frequency.%
}
\end{figure}

Since EM beams that are physically encoded with angular momentum
have a particular, azimuthally dependent phase behaviour (see
Table~\ref{tab:g_and_h}), it is important to investigate what impact
this physical fact might have, if any, on the possibilities to use
OAM in radio communications based on state-of-the art digital phase
modulation protocols. For this purpose we performed an outdoor OAM radio
experiment in an urban setting to test whether radio beams
carrying OAM can readily transfer digitally phase-modulated information
to a remote receiver under realistic conditions.

In the experiment described here, performed in Forte Marghera, Venice,
Italy, 2 August 2012, we used two collinear, linearly polarized radio
beams, one twisted ($m=1$) and one untwisted ($m=0$). As shown in
Fig.~\ref{fig:setup}, we generated the twisted beam with a helicoidally
deformed parabolic antenna, whereas the untwisted beam was generated
with a Yagi-Uda antenna, designed for the UHF S-band carrier frequency
used. Because of the different signatures of the respective phase
fronts of the two beams, the field vectors of a linearly polarized $m=1$
EM beam should be in anti-phase at two points on diametrically
opposite sides of the OAM phase singularity at the centre of the two
aligned beams, whereas they should be in phase for an $m=0$ beam (see
Fig.~\ref{fig:setup} and Ref.~\onlinecite[Fig.~2]{Thide&al:PRL:2007}).
We therefore probed the phases of the fields of the received signals
at such points in a standard phase interferometric manner, using
two identical antennas sensitive to the linear momentum carried by
the EM field. The measurements confirmed that the phases of the
received signals had the expected characteristics. This allowed us to
unambiguously identify and discriminate between the $m=0$ and $m=1$ OAM
eigenstates at the receiving end.

The transmitted $m=1$ signal was encoded with a DVB-S protocol
quadrature phase shift coding (QPSK) modulation. QPSK is a digital
phase encoding technique used in many telecommunications applications
today. It employs, at any given time, four different phase states
$\{01,11,10,00\}$ for the carrier. These four phase states correspond to
$\{0,90,180,270\}$ degrees of relative phase shifts, respectively. For
each temporal period, the phase can change once, while the amplitude
remains constant. In this way, two bits of information are conveyed
within each time slot. On the same UHF S-band carrier frequency,
we superimposed an untwisted ($m=0$) $100$~mW analogue frequency
modulation (FM) transmission with the same horizontal polarization
state as the OAM transmission. Both transmissions suffered reflections
from the ground as the beams propagated from the transmitters to the
$65$~m ($520\lambda$) distant receivers, both placed about $1.5$~m
($12\lambda$) above the reflecting ground. The QPSK constellation
diagram in Fig.~\ref{fig:QPSK_measured} was measured for information
transfer in a $17$~MHz bandwidth around the carrier frequency of
$2.414$~GHz in the presence of ground reflections and an interfering
$100$~MW FM signal, modulated with a TV test pattern, at the same
carrier frequency. As can be seen, the observed QPSK constellation
diagram agrees very well with the numerically simulated one shown in
Fig.~\ref{fig:QPSK_simulated}.

The modulation error ratio (MER) of the QPSK alone was larger than
$20$~dB, with a bit error rate (BER) of $10^{-8}$ and a carrier-to-noise
ratio C/N $>15$~dB, including the effect of ground reflections.

\subsection{Fading and reflections}

Fading caused by reflections is one of the most common complications
in wireless communications. Walls, ground and other objects can have a
detrimental effect on a communication channel and decrease the quality
of the transmission. On the other hand, reflections can also be utilized
in multi-path linear-momentum communication protocols such as MIMO
to increase the signal to noise radio and to enhance the information
transfer capability. So far we have not utilized any MIMO or MIMO-like
techniques in our OAM radio experiments. Such experiments are pending.

With OAM states, the problem is more intricate and subtle because
each reflection will introduce a parity change and OAM channel
swapping from left- to right-handed twist and \emph{vice versa}; see
Table~\ref{tab:g_and_h}. For the purpose of assessing the robustness of
the information transfer against perturbations, we used the reflection
of the waves off the ground with horizontal polarization, since this
maximizes the fading.

At the phase interferometer, the horizontally polarized OAM-carrying
electromagnetic beam was received as a superposition of the $m=1$ direct
beam and the reflected beam which, because of parity inversion, was
$m=-1$ charged. In order to assess the stability of the OAM mode, we
gradually varied the amount of disturbance introduced by reflections by
varying the inclination of the parabolic antenna transmitting the $m=1$
mode. This allowed us to superimpose the main twisted beam and
reflected beams with a reflected/transmitted beam ratio ranging from
$0.25$ to $0.5$ of the width of the receiving beam.

Because of the fading so produced, we measured a variation of the QPSK
signal from $9$ to $11$~dB and a resulting C/N ratio ranging from
$9$ to $12$~dB. The MER varied in the range $10$--$12$~dB and this
guaranteed an acceptable reception of the signal. 

To remove the problem of parity change, the receiver must be able
to discriminate between clockwise and counterclockwise vorticities
(positive and negative topological charge $m$). This can be achieved
by using either a selective phase mask or a circular array of $2m+1$
antennas \cite{Thide&al:PRL:2007}.

When an FM analogue transmitter signal with the same carrier frequency
was switched on, the ensuing interference on the digital, twisted signal
caused the MER to vary between $10$ and $12$~dB, the BER from
$10^{-3}$ to $10^{-5}$, and the C/N from $9$ to $15$~dB, for a variation
of reflected signal/signal ratio from $0.25$ to $0.5$.

\subsection{Summary}

Our results show that radio transmissions with OAM states are compatible
and robust with respect to digital multiplexing techniques, even those
based on phase coding such as phase shift keying (PSK). This is true
also when the OAM signal is disturbed by the presence of a strong
wide-band interfering signal on the same carrier frequency and by the
presence of ground reflections. The importance of our findings lies in
the fact that PSK protocols are at the core of the digital modulation
techniques used in modern telecommunications and broadcasting and in
many other of today's wireless scenarios. This offers the convenience
of back-compatibility between the new angular momentum and current
linear momentum radio methods.

Our experimental results are in full agreement with numerical
simulations performed. The maximum separation was $30$~dB whereas for
vertical polarization we estimate it to be up to $50$~dB. This clearly
shows that OAM can be used to increase the transmission capacity of our
common-use devices, allowing multiple services and users to share the
same frequency band.

We consider our experimental verification of the feasibility of using
OAM radio in communications applications using phase modulation a
significant leap forward, and a pivotal step toward the implementation
of novel radio concepts, applications and protocols.

\section{Methods}
\small


The digital transmitter used for the twisted ($m=1$) mode, a Microwave
Link QPSK DVB-S transmitter for the $2.4$~GHz band, was tuned to
$2.414$~GHz (free-space wavelength $\lambda=12.49$~cm). It transmitted
live encoded video images at a rate of $11.5$~Megasymbols/s. For
correction purposes, we used a forward error correction of the FEC=$3/4$
type, meaning that after three bits transmitted, a fourth bit was added.
This transmitter was connected to an $80$~cm ($6.4\lambda$) diameter
twisted parabolic antenna with a four elements patch feeder producing a
twisted ($m=1$) OAM beam.

In addition, an $1$~W analogue FM transmitter, fit with a $10$~dB
attenuator on the output, was used for transmitting a colour-bar
TV test pattern on the same frequency and along the same path. The
antenna used for this transmitter was a commercial-off-the-shelf (COTS)
$16$--$20$~dBi Yagi-Uda antenna, producing an untwisted ($m=0$) beam.

The two receiving antennas, used in a conventional phase interferometric
setup to measure the phase of the EM field, were two identical $26$~cm
($2.1\lambda$) diameter, $16$~dBi backfire antennas connected together
trough a signal splitter/combiner. A phase tuner, in the form of a
silver slit waveguide coupled to a Selenia signal circulator, was
inserted into one of the interferometer arms. By moving the cursor in
the slit waveguide, we retarded the signal received by one of the two
interferometer antennas relative to the other.

The signals collected by the interferometer were split up into three
different receiver chains: (1) a digital DVB-S digital chain with a
QPSK constellation tester, (2) a spectrum analyzer chain for measuring
and testing, and (3) an analogue frequency modulation (FM) chain. In
the latter chain we inserted a high-pass filter to block out the direct
current (DC) and audio frequency components.

The received signal was routed to an analyzer and/or to a 3-way
power splitter, where it was down-converted with a local oscillator
(LO) of $900$~MHz to $1.514$~GHz and split into three different
lines: (1) DVB-S, (2) analogue FM, and (3) the analyzer again.

The average background noise power in a $100$~MHz bandwidth was measured
at $-93$~dBm, peaking at $-85.6$~dBm at the centre frequency. The
power of the received FM signal was measured by inserting the spectrum
analyzer in the reception line and was found to be $-68.95$~dBm in a
$17$~MHz wide transmission band.


\bibliographystyle{naturemag}
\bibliography{bt,btlibrary,oam}

\section*{Acknowledgments}
 We gratefully acknowledge Andrea Bonifacio and Marco Polo System for the
 help and permission to use the experiment location Forte Marghera,
 and Tamara Vidali and Enzo Bon for their help and support during the
 experiment. B\,.T. was financially supported by the Swedish National
 Space Board (SNSB).

\end{document}